


\def   \ni {\noindent}
\def   \cl {\centerline}

\def   \ssk {\vskip  5truept}

\def   \bsk {\vskip 15truept}

\def   \newline {\hfil\break}


\magnification=1000

\hsize 5truein

\vsize 8truein

\font\abstract=cmr8

\font\text=cmr10

\font\affiliation=cmssi10

\font\author=cmss10

\font\title=cmssbx10 scaled\magstep2

\def\ref{\par\noindent\hangindent 15pt}

\nopagenumbers

\null

\vskip 3.0truecm

\baselineskip = 12pt

{\title

A physical distance indicator for spiral galaxies

\ni

}

\bsk \bsk

{\author

M. A. Hendry$^1$, S. Rauzy$^2$, P. Salucci$^3$, M. Persic$^{3,4}$

\ni

}

\bsk

{\affiliation

1. Astronomy Centre, University of Sussex, Brighton, BN1 9QH, UK

2. C.P.T. - C.N.R.S., Luminy Case 907, F-13288, Marseille, France

3. SISSA, Strada Costiera, 1-34014, Trieste, Italy

4. Osservatorio Astronomico, via G.B. Tiepolo 11, 1-34131, Trieste, Italy

}

\bsk

\cl {\it (Received )}

\bsk

\baselineskip = 9pt

{\abstract

In this paper we derive a Tully Fisher relation from measured I band
photometry and H$\alpha$ rotation curves of a large survey of southern sky
spiral galaxies, obtained in Persic \& Salucci (1995) by deprojecting and
folding the raw H$\alpha$ data of Mathewson, Ford \& Buchhorn (1992). We
calibrate the relation by combining several of the largest clusters in the
survey, using an iterative maximum likelihood procedure to account for
observational selection effects and Malmquist bias. We also incorporate a
simple model for the line of sight depth of each cluster. Our results
indicate a Tully Fisher relation of intrinsic dispersion $\sim0.3$ mag,
corresponding to a distance error dispersion of $13\%$. Application of this
relation to mapping the large scale velocity field is underway.

\ni

}

\bsk

\baselineskip = 12pt

{\text

\ni 1. INTRODUCTION

In recent years substantial progress has been made by Persic, Salucci and
collaborators in improving the physical basis of the Tully Fisher (TF)
relation. Their `mass decomposition' procedure -- when applied
to a sample of spirals with good quality rotation curves and B and V band
magnitudes --
yielded a TF relation which was more linear and displayed less scatter than
its uncorrected counterpart (Salucci, Frenk \& Persic, 1993).
In this work a similar attempt is made to improve the calibration of
the TF relation based on the I band CCD photometry and H$\alpha$ rotation
curves of a sample of cluster spirals from the southern sky redfshift survey of
Mathewson, Ford \& Buchhorn (1992; hereafter MFB). This provides a larger
and more accurate database than that considered in Salucci, Frenk \& Persic
(1993), and avoids the use of redshift as a direct measure of
distance. In addition, we also carefully
address here the issues of selection bias and undersampling, which it has
been suggested (c.f. Sandage, Tammann \& Federspiel, 1995; hereafter STF)
have been inadequately dealt with in the cluster calibration of MFB.

\bsk

\ni 2. THE TULLY FISHER CALIBRATION DATA

The re-folding and smoothing of the raw MFB data is described fully in
Persic \& Salucci (1995). For each galaxy an optical radius, $R_{\rm{opt}}$,
encompassing 83\% of the integrated light, was then
computed and a smoothed rotation velocity obtained at a series of standard
radii -- each corresponding to a fixed fraction of $R_{\rm{opt}}$. The
calibrating sample was then selected from the subset of 161 galaxies
identified in MFB as belonging to clusters or groups.

We used the smoothed rotation velocity at a radius of $0.6R_{\rm{opt}}$
(denoted by $V_6$) in the calibration: at this radius the contribution to
the rotation velocity from the luminous disk is a maximum in the disk model
of Salucci, Frenk \& Persic (1993) and earlier papers -- thus leading to a TF
relation with a stronger physical basis than that obtained in MFB, where the
maximum measured rotation velocity was used. These velocities were
then combined with I band magnitudes, integrated to a radius of
$0.6R_{\rm{opt}}$ (c.f. Stel, 1994; denoted here by $I_6$), and estimates of
$\Delta$, the ratio of semimajor and semiminor axis, to derive a linear TF
relation.

\bsk

\ni 3. MAXIMUM LIKELIHOOD METHOD

The MFB TF calibration used a sample of 14 galaxies in the Fornax cluster,
for which a very strong correlation was obtained. This calibration has been
criticised for failing to include the effects of selection bias and
misrepresenting the {\it true\/} slope and dispersion of the relation
due to the small sample size (Hendry \& Simmons, 1994, hereafter
HS; STF). In HS it was shown that for a
calibrating sample of $\sim30$ galaxies or less, the sampling distribution of
the TF slope typically has a {\it larger\/} dispersion than the bias
due to selection effects, so that sampling error -- not selection bias
-- is the dominant systematic uncertainty. In this work we address
{\it both\/} problems, however. We take account of luminosity
selection following the statistical formalism of Hendry \& Simmons (1990)
and HS. We also considerably increase the calibrating sample size
by combining data from several MFB clusters.

The full details of our analysis will be described in Hendry et al (1995),
and we merely summarise the main points here. We assume that the
{\it intrinsic\/} conditional distribution of absolute magnitude, $M_I$,
given $\log V_6$ and $\log \Delta$, is a normal distribution with dispersion
$\sigma$ and mean value a linear function of $\log V_6$ and $\log \Delta$,
i.e.
$$
E ( M_I | \log V_6 , \log \Delta ) \quad = \quad
\alpha \log V_6 + \beta \log \Delta + \gamma
$$
We {\it impose\/} a sharp selection limit at apparent magnitude, $I_{\rm{L}}$
(c.f. STF), and thus derive the conditional distribution of $M_I$ given
$\log V_6$ and $\log \Delta$ for an {\it observable\/} galaxy at true
distance, $r$.

Suppose the calibrating sample consists of $n$ different clusters,
assumed (for now) all to lie at the same distance. For the $j^{th}$ galaxy:-
$$
M_I^j \quad = \quad I_6^j - \sum_{k=1}^{n} z_{jk} 5 \log r_k - 25
$$
where $r_k$ is the distance (in Mpc) of the $k^{th}$ cluster and
$z_{jk} = 1$ if the $j^{th}$ galaxy belongs to the $k^{th}$ cluster
and $z_{jk} = 0$ otherwise. We substitute this expression into the
likelihood function and thus obtain maximum likelihood (ML) estimates for
$\alpha$, $\beta$, $\gamma$, the $r_k$ and $\sigma$. Although no closed
analytic form exists for the ML solution, the equations are easily solved
iteratively. Following MFB, we assume a redshift distance of 1340kms$^{-1}$
for Fornax and take $H_0=50$, although these choices have no bearing on the
ML estimate of $\sigma$.

\bsk

\ni 4. RESULTS

We selected first the subsample of 68 spirals with $I_6 \leq 15$, from the
six largest MFB clusters: Antlia, Eridanus, Fornax, Hydra, Pegasus and
Sculptor. Applying the ML method yielded an estimate of $\hat{\sigma} =
0.41 \pm 0.04$. (Standard errors were calculated from Monte Carlo
simulations). We then investigated the effect of selecting only a subset of
the clusters. Clearly the optimal solution is one for which the decrease in
sample size by eliminating one or more clusters is balanced by the decrease
in the number of free parameters (c.f. Hendry et al, 1995). Selecting the 50
galaxies in Fornax, Hydra, Antlia and Pegasus yielded $\hat{\sigma} = 0.33
\pm 0.04$ -- a significantly smaller dispersion. Adding Sculptor to these
four clusters
gave a sample size of 59 and $\hat{\sigma} = 0.35 \pm 0.03$. Including
Eridanus instead of Sculptor, on the other hand, gave
$\hat{\sigma} = 0.41 \pm 0.04$, with similar ML estimates of $\alpha$,
$\beta$ and $\gamma$ to those of the six-cluster fit.
That a problem existed with Eridanus was apparent from the value
of $\hat{\sigma} = 0.70 \pm 0.08$ obtained when fitting the TF relation to
Eridanus alone -- almost twice as large as the next largest estimate for an
individual cluster. The most likely reason for this large dispersion was that
the galaxies did not all lie at the same true distance -- a conclusion
strongly supported when discarding only 3 galaxies from Eridanus
reduced $\hat{\sigma}$ to $0.33 \pm 0.06$.

The `canonical' ML fit obtained from Antlia, Fornax, Hydra and Pegasus
was robust when restricted to various subsets of these clusters  -- although
further reduction in the sample size increased the formal errors on the
fitted parameters. We therefore adopted this solution as our optimal TF
relation, viz.
$$
M_I \quad = \quad (-6.04 \pm 0.31) \log V_6 + (0.66 \pm 0.25) \log \Delta
+ (-8.19 \pm 0.66)
$$

The discovery of significant line of sight depth in Eridanus
led us to consider next to what extent our ML value of
$\hat{\sigma} = 0.33 \pm 0.04$ might be slightly overestimated due to line of
sight depth.

\bsk

\ni 5. THE EFFECT OF LINE OF SIGHT DEPTH ON $\sigma$

We modelled the effect of line of sight depth in each cluster by assuming
the galaxies' distance modulus to be normally distributed about some mean
value, $\mu_{\rm{clus}}$, with dispersion, $\sigma_{\rm{clus}}$ -- an
assumption borne out of mathematical expediency, but one which numerical
simulations show to have negligible bearing on our final results
(c.f. Hendry et al, 1995). In this case the dispersion in distance
modulus adds quadratically to the intrinsic scatter of the TF relation.
Thus, one may determine a corrected estimate of $\sigma$ according to:-
$$
\hat{\sigma}^2_{\rm{corr}} \quad = \quad \hat{\sigma}^2_{\rm{obs}} -
\sigma^2_{\rm{clus}}
$$
We estimated $\sigma_{\rm{clus}}$ from the projected angular dispersion,
assuming galaxies to be isotropically distributed about the cluster
centre. This assumption is likely to be unreasonable for
an individual cluster but one might expect it to hold statistically when we
merge together several clusters. From our `canonical' sample we thus
obtained $\sigma_{\rm{clus}} = 0.15$, and hence derived a corrected estimate
for the intrinsic dispersion of the TF relation of
$$
\hat{\sigma}_{\rm{corr}} \quad = \quad 0.30 \pm 0.04
$$

\bsk

\ni 6. SUMMARY

In this work we have derived an optimal TF relation using a
calibrating sample of spiral galaxies from the southern sky redshift survey
originally published in MFB. We have improved the physical basis of the
relation by deriving I band apparent magnitudes and rotation velocities at
a fixed fraction of the optical radius for each galaxy -- specifically at
the radius for which the contribution to the rotation velocity from the
luminous matter is maximised for a Freeman disk model.

We have accounted for the effects of luminosity selection
by applying a maximum likelihood analysis to a composite sample of four
clusters, determing optimal estimates for the TF coefficients and the
relative cluster distances in a self-consistent manner. We derive a TF
dispersion of $\hat{\sigma} = 0.33 \pm 0.04$, which we strongly believe
does not
underestimate the {\it true\/} TF scatter since our calibrating sample is
sufficiently large to overcome criticisms of undersampling.

We have identified a likely source of significant line of sight depth in the
Eridanus cluster -- a group of three galaxies whose removal reduces the
TF dispersion in Eridanus to a value which is completely consistent with the
remaining clusters and with our composite relation.

Finally, we have modelled the contribution to the observed TF scatter from
residual line of sight depth using the projected distribution of galaxy
positions, assuming that our composite cluster sample should be
statistically isotropic -- an assumption which would be considerably less
justifiable for an individual cluster. This hypothesis leads to a corrected
TF dispersion of $\hat{\sigma} = 0.30 \pm 0.04$, which we adopt as our
estimate
of the true dispersion of our best-fit TF relation. This corresponds to a
distance error dispersion of $13\%$.

We are currently applying this relation to derive TF distance estimates to the
remaining clusters and field galaxies of Persic \& Salucci (1995), and are
using these as input to various methods for reconstructing the large
scale velocity and density fields.

\bsk

\vskip 0.1truecm

\ni {REFERENCES}

\ssk

\ref{Hendry, M.A., Simmons, J.F.L., 1990, A\&A, 237, 275}
\ref{Hendry, M.A., Simmons, J.F.L., 1994, ApJ, 435, 515 (HS)}
\ref{Hendry, M.A., Rauzy, S., Salucci, P., Persic, M., 1995, in prep.}
\ref{Mathewson D.S., Ford V.L., Buchhorn M., 1992, ApJS, 81, 413 (MFB)}
\ref{Persic M., Salucci P., 1995, ApJS, in press}
\ref{Salucci P., Frenk C.S., Persic M., 1993, MNRAS, 262, 392}
\ref{Sandage A., Tammann G.S., Federspiel M., 1995, preprint (STF)}
\ref{Stel F., 1994, Laurea Degree Thesis, University of Trieste}

}

\end